\documentclass[twocolumn,showpacs,preprintnumbers,amsmath,amssymb]{revtex4}

\usepackage{graphicx}
\usepackage{dcolumn}
\usepackage{bm}

\begin{document}

\preprint{}
\input{epsf.tex}

\epsfverbosetrue

\title{Collective Light Emission of a Finite Size Atomic Chain}

\author{Hashem Zoubi}

\affiliation{Institut f$\ddot{u}$r Theoretische Physik, Universit$\ddot{a}$t Innsbruck, Technikerstrasse 25, A-6020 Innsbruck, Austria}

\date{09 March, 2012}

\begin{abstract}
Radiative properties of collective electronic states in a one dimensional atomic chain are investigated. Radiative corrections are included with emphasize put on the effect of the chain size through the dependence on both the number of atoms and the lattice constant. The damping rates of collective states are calculated in considering radiative effects for different values of the lattice constant relative to the atomic transition wave length. Especially the symmetric state damping rate as a function of the number of the atoms is derived. The emission pattern off a finite linear chain is also presented. The results can be adopted for any chain of active material, e.g., a chain of semiconductor quantum dots or organic molecules on a linear matrix.
\end{abstract}


\maketitle

\section{Introduction}

Optical lattice ultracold atoms continue to be of interest for more and more researches of different branches of physics \cite{Dalibard}. Big attention is given for the realization of different condensed matter models that provide a test system for achieving a deep understanding of fundamental physics and answering open questions in the subject \cite{Lewenstein}, beside their applications for quantum information processing \cite{Zeilinger}. In general, the main objective is to consider optical lattice ultracold atoms as artificial crystals with a wide range of controllable parameters.

Optical lattices form of counter propagating laser beams to get standing waves in which ground state ultracold atoms are loaded \cite{Bloch,Jaksch}. The atoms experience optical lattice potential with lattice constant of half wave length of the laser. Low dimensional lattices can be achieved with different geometric structures and symmetries \cite{Spielman}. In conventional solid crystals the lattice constant and the symmetry of the lattice is fixed through the different chemical bonds that responsible for the formation of the crystal. The advantage of optical lattices is due to the controllability of the lattice constant and symmetry through controlling the external laser field \cite{Dalibard}.

Collective states of electronic excitations play a central rule in solid crystals and molecular clusters and they usually termed excitons \cite{Davydov,Agranovich}. They induced by electrostatic interactions among the lattice atoms or molecules, where an electronic excitation can be delocalized in the crystal through energy transfer. Collective states can dominate the electrical and optical properties of the material, and especially they strongly affect the excitation lifetimes and give rise to dark and superradiant states. In such material the lattice constant is few angstroms which is much smaller than the electronic transition wavelength, and hence one can use electrostatic interactions, e.g. resonance dipole-dipole interactions, and to neglect radiative corrections altogether.

Electronic excitations in optical lattice ultracold atoms are of big importance, e.g., for optical lattice clocks \cite{Katori}, and for optical lattice Rydberg atoms \cite{Arimondo}. In our previous work we introduced excitons for optical lattice ultracold atoms in one and two dimensional set-ups \cite{ZoubiA,ZoubiB}. We concentrated mainly in the Mott insulator phase with one and two atoms per lattice site. We treated both large and finite atomic chains \cite{ZoubiC,ZoubiD,ZoubiE}, and we calculated the damping rate of excitons into free space and their emission pattern \cite{ZoubiF,ZoubiG,ZoubiH}. In all of our previous researches we exploited electrostatic interactions for the formation of collective states, mainly resonance dipole-dipole interactions. But for typical optical lattices the lattice constant is few thousands of angstroms, which can be of the order of the electronic transition wavelength, and hence radiative corrections can be significant.

In the present paper we investigate a one dimensional finite chain of atoms where the lattice constant can take any value relative to the atomic transition wavelength. Finite atomic chains have been realized recently in a number of optical lattice experiments \cite{Vetsch,Weitenberg}. We emphasize the influence of radiative corrections on the formation of collective sates and their damping rates, where we exploit general collective states with emphasize on the most symmetric one. We derive the condition for the validity of applying electrostatic interactions, which we used in our previous work. Few studies treated the collective effect on the optical properties of finite atomic chain of several atoms \cite{Mewton}, but extensive study done for two atoms in the radiative regime \cite{Ficek}, and in which we compare our results. We extract how the damping rate depends on the chain size, namely on the number of atoms in the lattice. Furthermore, we calculate the emission pattern off a finite atomic chain.

The paper is organized as follows: in section 2 we present a finite one dimensional atomic chain and discuss the energy transfer parameter due to dipole-dipole interactions in the radiative regime. Then in section 3 we calculate the damping rates for different collective states and several chain sizes. The emission pattern for collective states is calculated in section 4. The summary appears in section 5.

\section{Finite One-Dimensional Atomic Chain}

We consider a finite one dimensional atomic lattice, where the number of atoms is $N$ with lattice constant $a$, as seen in figure (1). The atoms are considered to be two-level systems with electronic transition energy $E_A=\hbar\omega_A$. An electronic excitation can delocalize in the lattice by transferring among the atoms. The electronic excitation Hamiltonian is given by
\begin{equation}
H_{ex}=\sum_n\hbar\omega_A\ B_n^{\dagger}B_n+\sum_{nm}\ \hbar J_{nm}B_n^{\dagger}B_m,
\end{equation}
where $B_n^{\dagger}$ and $B_n$ are the creation and annihilation operators of an electronic excitation at atom $n$. For a single excitation the operators can be assumed to obey boson commutation relations. 

\begin{figure}[h!]
\centerline{\epsfxsize=7cm \epsfbox{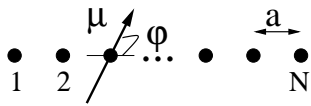}}
\caption{A finite lattice of $N$ atoms. The lattice constant is $a$, and the transition dipole $\mbox{\boldmath$\mu$}$ makes an angle $\varphi$ with the lattice direction.}
\end{figure}

The energy transfer among two atoms, $n$ and $m$, is a function of the interatomic distance and given by \cite{Craig}
\begin{eqnarray}\label{Exact}
J\left(q_AR_{nm}\right)&=&\frac{3}{4}\Gamma_A\left\{\left[\frac{\sin\left(q_AR_{nm}\right)}{\left(q_AR_{nm}\right)^2}+\frac{\cos\left(q_AR_{nm}\right)}{\left(q_AR_{nm}\right)^3}\right]\right. \nonumber \\
&\times&\left.\left(1-3\cos^2\varphi\right)\right. \nonumber \\
&-&\left.\frac{\cos\left(q_AR_{nm}\right)}{q_AR_{nm}}\left(1-\cos^2\varphi\right)\right\},
\end{eqnarray}
where the distance between the two atoms is $R_{nm}=|n-m|a$, and $\mu$ is the magnitude of the electronic excitation transition dipole, which makes an angle $\varphi$ with the lattice direction, see figure (1). $q_A$ is the atomic transition wave number given by $E_A=\hbar cq_A$. Here $\Gamma_A$ is the single excited atom damping rate
\begin{equation}
\Gamma_A=\frac{\omega^3_A\mu^2}{3\pi\epsilon_0\hbar c^3}.
\end{equation}
In the limit of $\lambda_A>a$, where $\lambda_A$ is the atomic transition wave length defined by $E_A=hc/\lambda_A$, we can consider only energy transfer among nearest neighbor atoms with $J\left(q_Aa\right)$ where we take $R_{nm}=a$.

In figure (2) we plot $J\left(q_Aa\right)/\Gamma_A$ as a function of $q_Aa$ for two different polarization directions. Note that for typical optical lattice we have $E_A=1\ eV$, with $\lambda_A\approx 12405\ \AA$, and $q_A\approx 4\times10^{-4}\ \AA^{-1}$. For $a=1000\ \AA$ we get $q_Aa\approx 0.5$, and $a/\lambda_A\approx 0.08$. For $\varphi=0^{\circ}$ we obtain $J(0.5)/\Gamma_A\approx -13.4$, and for $\varphi=90^{\circ}$ we get $J(0.5)/\Gamma_A\approx 5.4$. For large $q_Aa$ the coupling tend to zero with oscillations, and the atoms are almost independent.

\begin{figure}[h!]
\centerline{\epsfxsize=8cm \epsfbox{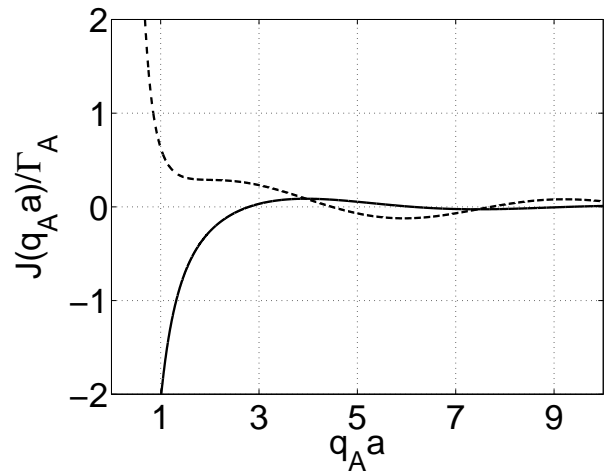}}
\caption{The scaled interaction $J\left(q_Aa\right)/\Gamma_A$ vs. $q_Aa$. The full line is for $\varphi=0^{\circ}$, and the dashed line for $\varphi=90^{\circ}$.}
\end{figure}

In the limit $\lambda_A\gg a$, or $q_Aa\ll1$, we can neglect the radiative terms (as we did in our previous works \cite{ZoubiA,ZoubiB,ZoubiC,ZoubiD,ZoubiE,ZoubiF,ZoubiG,ZoubiH}), to get the electrostatic resonance dipole-dipole interaction
\begin{equation}\label{Appr}
J\approx\frac{3}{4}\frac{\Gamma_A}{(q_Aa)^3}\left(1-3\cos^2\varphi\right).
\end{equation}
Using the previous numbers, $\varphi=0^{\circ}$ yields $J(0.5)/\Gamma_A\approx -12$, and $\varphi=90^{\circ}$ yields $J(0.5)/\Gamma_A\approx 6$, which are slightly different from the above exact results. For smaller $q_Aa$ we get much better agreement. In figure $(3)$ we plot equations (\ref{Exact}) and (\ref{Appr}) for $\varphi=0^{\circ}$, and in figure $(4)$ for $\varphi=90^{\circ}$. The results justify the use of electrostatic dipole-dipole interactions for optical lattice ultracold atoms when $q_Aa<1$.

\begin{figure}[h!]
\centerline{\epsfxsize=8cm \epsfbox{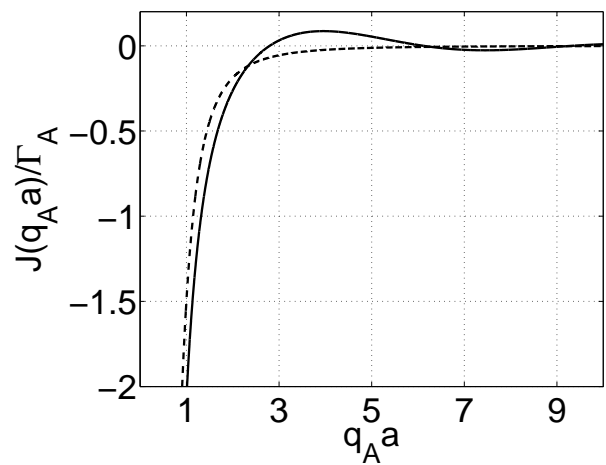}}
\caption{The scaled interaction $J\left(q_Aa\right)/\Gamma_A$ vs. $q_Aa$ for $\varphi=0^{\circ}$. The full line is for equation (\ref{Exact}), and the dashed line for equation (\ref{Appr}).}
\end{figure}

\begin{figure}[h!]
\centerline{\epsfxsize=8cm \epsfbox{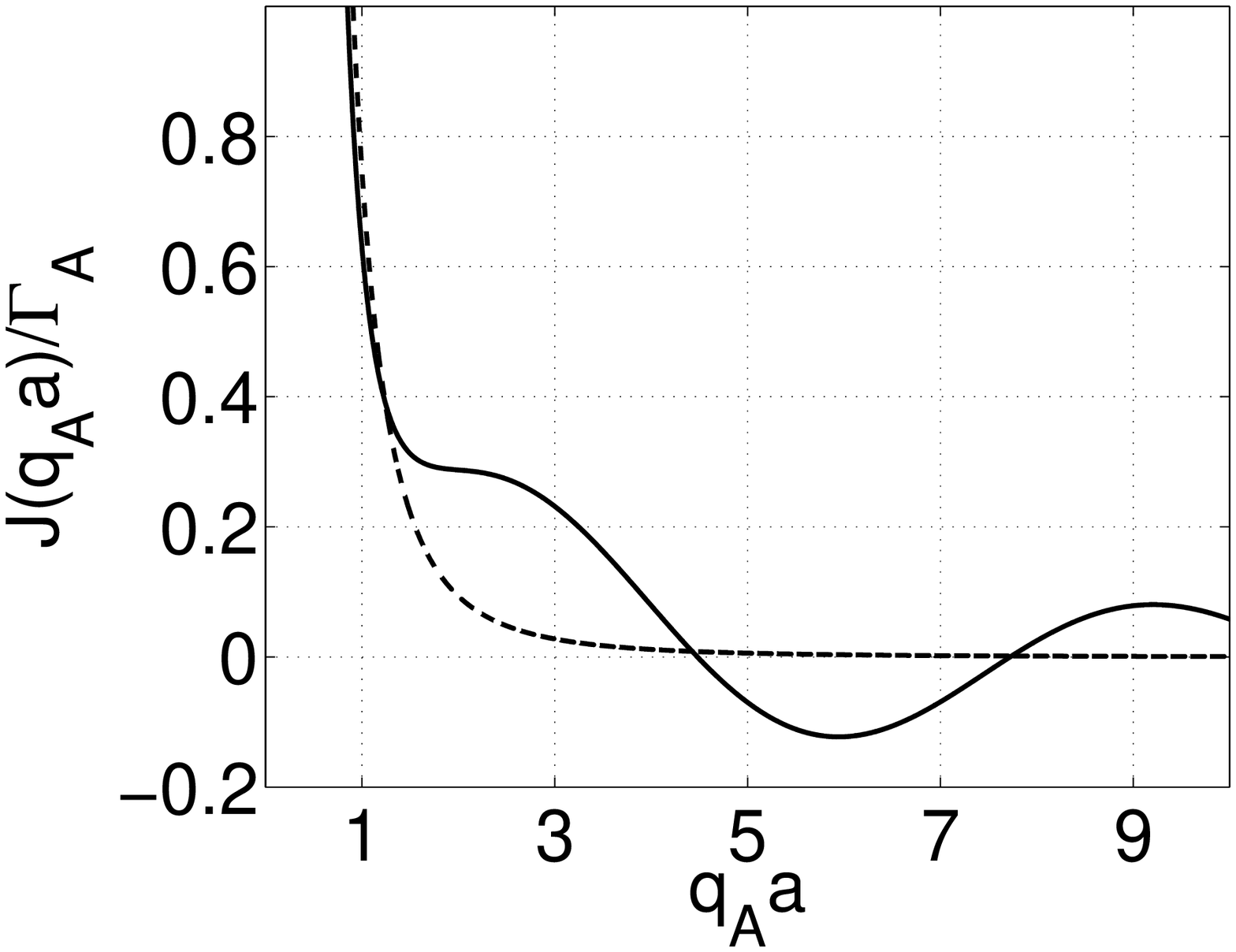}}
\caption{The scaled interaction $J\left(q_Aa\right)/\Gamma_A$ vs. $q_Aa$ for $\varphi=90^{\circ}$. The full line is for equation (\ref{Exact}), and the dashed line for equation (\ref{Appr}).}
\end{figure}

\section{Collective Excitation Damping Rate}

We start in presenting the free space radiation field and its coupling to a finite atomic chain. The free space radiation field Hamiltonian is
\begin{equation}
H_{rad}=\sum_{{\bf q}\lambda}E_{ph}(q)\ a_{{\bf q}\lambda}^{\dagger}a_{{\bf q}\lambda},
\end{equation}
where $a_{{\bf q}\lambda}^{\dagger}$ and $a_{{\bf q}\lambda}$ are the creation and annihilation operators of a photon with wave vector ${\bf q}$ and polarization $\lambda$, respectively. The photon energy is $E_{ph}(q)=\hbar cq$. The electric field operator is
\begin{equation}
\hat{\bf E}({\bf r})=i\sum_{{\bf q}\lambda}\sqrt{\frac{\hbar cq}{2\epsilon_0 V}}\left\{a_{{\bf q}\lambda}\ {\bf e}_{{\bf q}\lambda}e^{i{\bf q}\cdot{\bf r}}-a_{{\bf q}\lambda}^{\dagger}\ {\bf e}_{{\bf q}\lambda}^{\ast}e^{-i{\bf q}\cdot{\bf r}}\right\},
\end{equation}
where ${\bf e}_{{\bf q}\lambda}$ is the photon polarization unit vector, and $V$ is the normalization volume.

The atomic transition dipole operator is 
\begin{equation}
\hat{\mbox{\boldmath$\mu$}}=\mbox{\boldmath$\mu$}\sum_{n=1}^N\left(B_n+B_n^{\dagger}\right).
\end{equation}
The matter-field coupling is given formally by the electric dipole interaction $H_I=-\hat{\mbox{\boldmath$\mu$}}\cdot\hat{\bf E}$. In the rotating wave approximation and for linear polarization, we get
\begin{eqnarray}
H_I&=&-i\sum_{{\bf q}\lambda,n}\sqrt{\frac{\hbar cq}{2\epsilon_0 V}}\left(\mbox{\boldmath$\mu$}\cdot{\bf e}_{{\bf q}\lambda}\right) \nonumber \\
&\times&\left\{a_{{\bf q}\lambda}B_n^{\dagger}\ e^{iq_zna}-a_{{\bf q}\lambda}^{\dagger}B_n\ e^{-iq_zna}\right\}.
\end{eqnarray}
In the following we treat a single electronic excitation in the atomic chain. We start in treating the most symmetric collective state and then the general collective state.

\subsection{Symmetric Collective Excitation}

We consider a single excitation in the system with the symmetric collective state
\begin{equation}
|i\rangle_s=\frac{1}{\sqrt{N}}\sum_i|g_1,\cdots,e_i,\cdots,g_N\rangle.
\end{equation}
This state is an eigenstate of the Hamiltonian in the limit of $q_Aa>1$ with $J/\Gamma_A<1$, where the atoms are almost independent. The other limit of $q_Aa<1$ treated by us in other work \cite{ZoubiA,ZoubiB,ZoubiC,ZoubiD,ZoubiE,ZoubiF,ZoubiG,ZoubiH}.

We calculate the damping rate of such collective state through the emission of a photon into free space and the damping into the final ground state
\begin{equation}
|f\rangle=|g_1,\cdots,g_N\rangle.
\end{equation}
We apply the Fermi golden rule to calculate the collective symmetric state damping rate
\begin{equation}
\Gamma_s=\frac{2\pi}{\hbar}\sum_{{\bf q}\lambda}|\langle f|H_I|i\rangle|^2\delta(E_A-E_{ph}),
\end{equation}
which in the present case reads
\begin{equation}
\Gamma_s=\sum_{{\bf q}\lambda}\frac{\pi cq}{\epsilon_0 VN}\left(\mbox{\boldmath$\mu$}\cdot{\bf e}_{{\bf q}\lambda}\right)^2\left|\sum_{n=1}^Ne^{-iq_zna}\right|^2\delta(E_A-E_{ph}).
\end{equation}
The summation over the photon polarization yields
\begin{equation}
\sum_{\lambda}\left(\mbox{\boldmath$\mu$}\cdot{\bf e}_{{\bf q}\lambda}\right)^2=\mu^2-\frac{\left({\bf q}\cdot\mbox{\boldmath$\mu$}\right)^2}{q^2}.
\end{equation}
The summation over ${\bf q}$ can be converted into the integral
\begin{equation}
\sum_{\bf q}\rightarrow \frac{V}{(2\pi)^3}\int_{0}^{2\pi}d\phi\int_{0}^{\pi}d\theta\sin\theta\int_0^{\infty}q^2 dq.
\end{equation}
We use
\begin{equation}
{\bf q}=q(\sin\theta\cos\phi,\sin\theta\sin\phi,\cos\theta),
\end{equation}
and the transition dipole is taken to be
\begin{equation}
\mbox{\boldmath$\mu$}=\mu(\sin\varphi,0,\cos\varphi).
\end{equation}
The integration over $\phi$, and the change of the variable $y=q_Aa\ cos\theta$, gives
\begin{eqnarray}\label{Gamma}
\Gamma_s&=&\frac{\mu^2q_A^2}{8\pi\epsilon_0 \hbar aN}\int_{-q_Aa}^{+q_Aa}dy\left|\sum_{n=1}^Ne^{-iny}\right|^2 \nonumber \\
&\times&\left[\left(1+\cos^2\varphi\right)-\frac{y^2}{(q_Aa)^2}\left(3\cos^2\varphi-1\right)\right].
\end{eqnarray}
Using the relation
\begin{equation}
\left|\sum_{n=1}^Ne^{-iny}\right|^2=N+\sum_{n<m=1}^N2\cos[(n-m)y],
\end{equation}
we reach, after the integration over $y$, the result
\begin{equation}
\Gamma_s=\Gamma_A\left\{1+\frac{2}{N}\sum_{n<m=1}^NF[q_Aa(n-m)]\right\},
\end{equation}
where
\begin{eqnarray}
F(x)&=&\frac{3}{2}\left\{\frac{\sin x}{x}\left(1-\cos^2\varphi\right)\right. \nonumber \\
&+&\left.\left[\frac{\cos x}{x^2}-\frac{\sin x}{x^3}\right]\left(1-3\cos^2\varphi\right)\right\}.
\end{eqnarray}
In figure (5) we plot the function $F(x)$, for two different polarization directions. Using the previous numbers, for $\varphi=0^{\circ}$ we get $F(0.5)=0.9752$, and for $\varphi=90^{\circ}$ we get $0.9507$, which justifies the use of $F=1$ for optical lattice ultracold atoms in our previous works \cite{ZoubiA,ZoubiB,ZoubiC,ZoubiD,ZoubiE,ZoubiF,ZoubiG,ZoubiH}.

\begin{figure}[h!]
\centerline{\epsfxsize=8cm \epsfbox{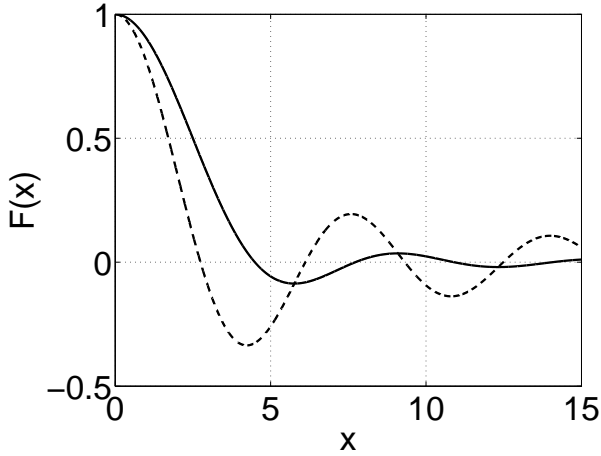}}
\caption{The function $F(x)$ vs. $x$. The full line is for $\varphi=0^{\circ}$, and the dashed line for $\varphi=90^{\circ}$.}
\end{figure}

For different number of atoms we get
\begin{eqnarray}
\Gamma_s(1)&=&\Gamma_A, \nonumber \\
\Gamma_s(2)&=&\Gamma_A\left\{1+F(q_Aa)\right\}, \nonumber \\
\Gamma_s(3)&=&\Gamma_A\left\{1+\frac{2}{3}\left[2F(q_Aa)+F(2q_Aa)\right]\right\}, \nonumber \\
&\cdots&
\end{eqnarray}
The symmetric damping rate can be written in the form
\begin{equation}
\Gamma_s(N)=\Gamma_A\left\{1+2\sum_{n=1}^{N-1}\frac{(N-n)}{N}F(q_Aan)\right\}.
\end{equation}
In figure $(6)$ we plot $\Gamma/\Gamma_A$ as a function of $q_Aa$ for $N=5$, and for the polarizations $\varphi=0^{\circ}$ and $\varphi=90^{\circ}$.

\begin{figure}[h!]
\centerline{\epsfxsize=8cm \epsfbox{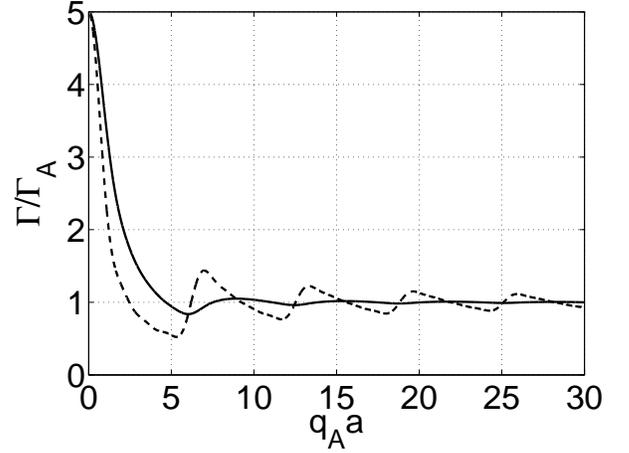}}
\caption{The symmetric scaled damping $\Gamma/\Gamma_A$ vs. $q_Aa$, for $N=5$. The full line is for $\varphi=0^{\circ}$, and the dashed line for $\varphi=90^{\circ}$.}
\end{figure}

Lets consider $F(q_Aan)$ to represent a bond between two atoms that separated by a distance $(an)$, then in the above summation the function $F(q_Aan)$ is multiplied by the number of bonds of this length which is $(N-n)$. In the limit of $q_Aa\ll1$ we get $F\simeq 1$, and then $\Gamma_s(N)\approx N\Gamma_A$. In the limit of $q_Aa\gg1$ we get $F\simeq 0$, and then $\Gamma_s(N)\approx\Gamma_A$ with oscillations.

Now we emphasize the dependence of the symmetric state damping rate as a function of the number of atoms $N$. We plot the scaled damping rate $\Gamma_s/\Gamma_A$ as a function of $N$ for different values of $q_Aa$. In figures $(7-9)$ we plot for $q_Aa=0.001$, $q_Aa=0.1$, and $q_Aa=1$, in the two cases of $\varphi=0^{\circ}$ and $\varphi=90^{\circ}$. The damping rate of the symmetric state grows linearly with the number of atoms for small $N$, and approach a finite value for large $N$. For $q_Aa\geq1$ the damping rate approach the finite value faster than for $q_Aa\ll1$. In figure $(10)$ we plot $\Gamma_s/\Gamma_A$ as a function of $\phi$ for $N=100$ at $q_Aa=0.1$. Significant difference appears between the damping rates for $\varphi=0^{\circ}$ and $\varphi=90^{\circ}$.

\begin{figure}[h!]
\centerline{\epsfxsize=8cm \epsfbox{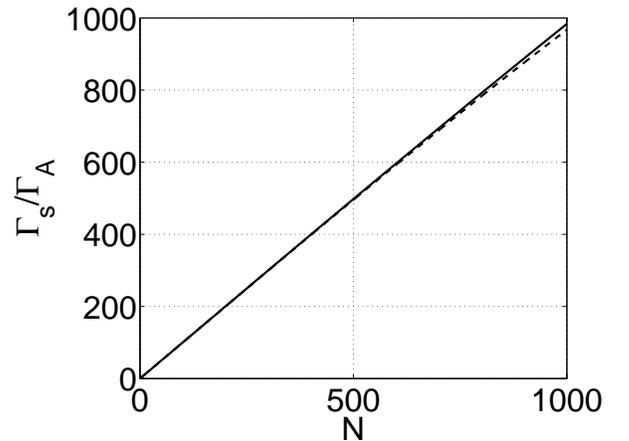}}
\caption{The symmetric scaled damping rate $\Gamma/\Gamma_A$ vs. $N$, for $q_Aa=0.001$. The full line is for $\varphi=0^{\circ}$, and the dashed line for $\varphi=90^{\circ}$.}
\end{figure}

\begin{figure}[h!]
\centerline{\epsfxsize=8cm \epsfbox{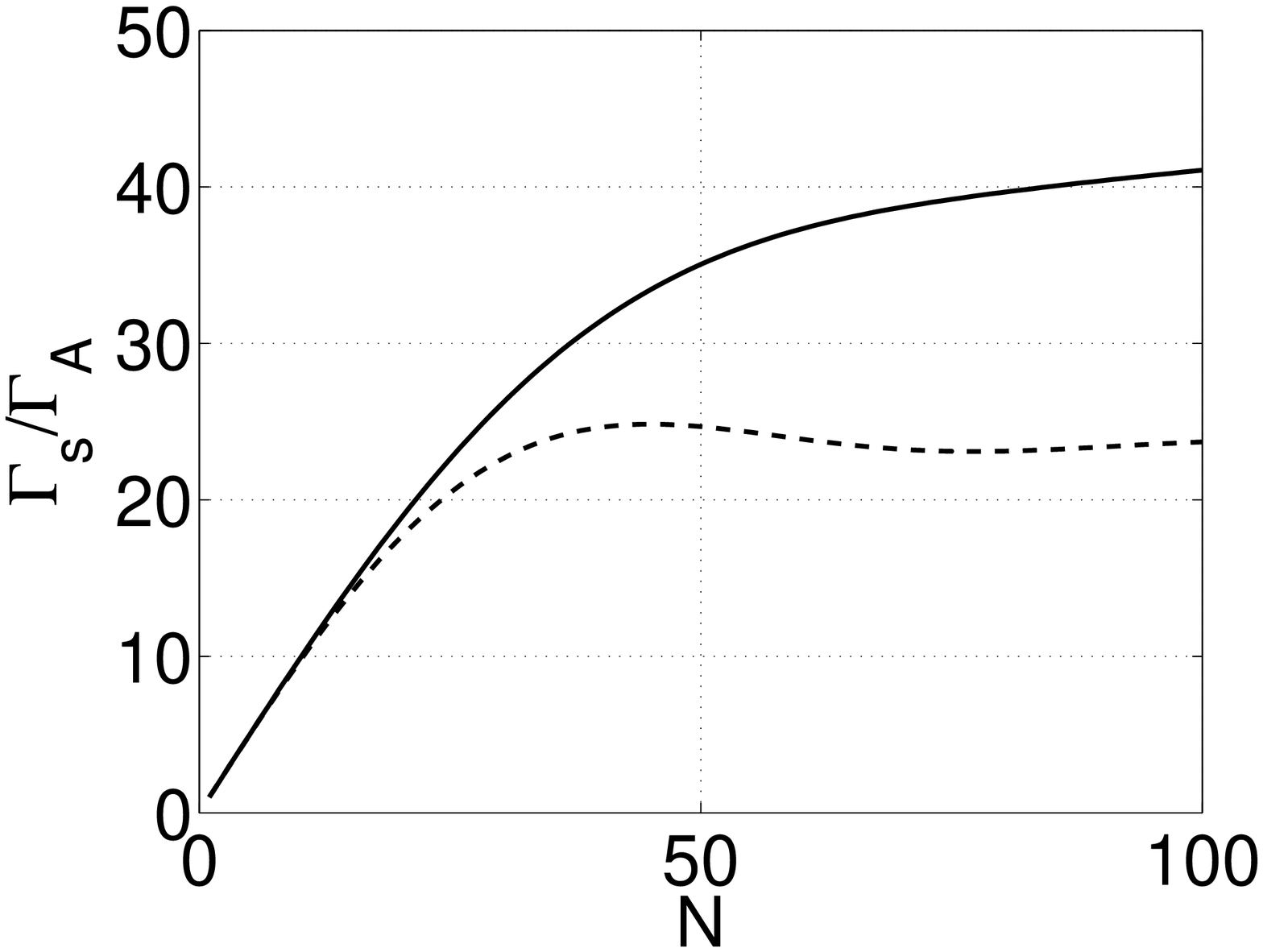}}
\caption{The symmetric scaled damping rate $\Gamma/\Gamma_A$ vs. $N$, for $q_Aa=0.1$. The full line is for $\varphi=0^{\circ}$, and the dashed line for $\varphi=90^{\circ}$.} 
\end{figure}

\begin{figure}[h!]
\centerline{\epsfxsize=8cm \epsfbox{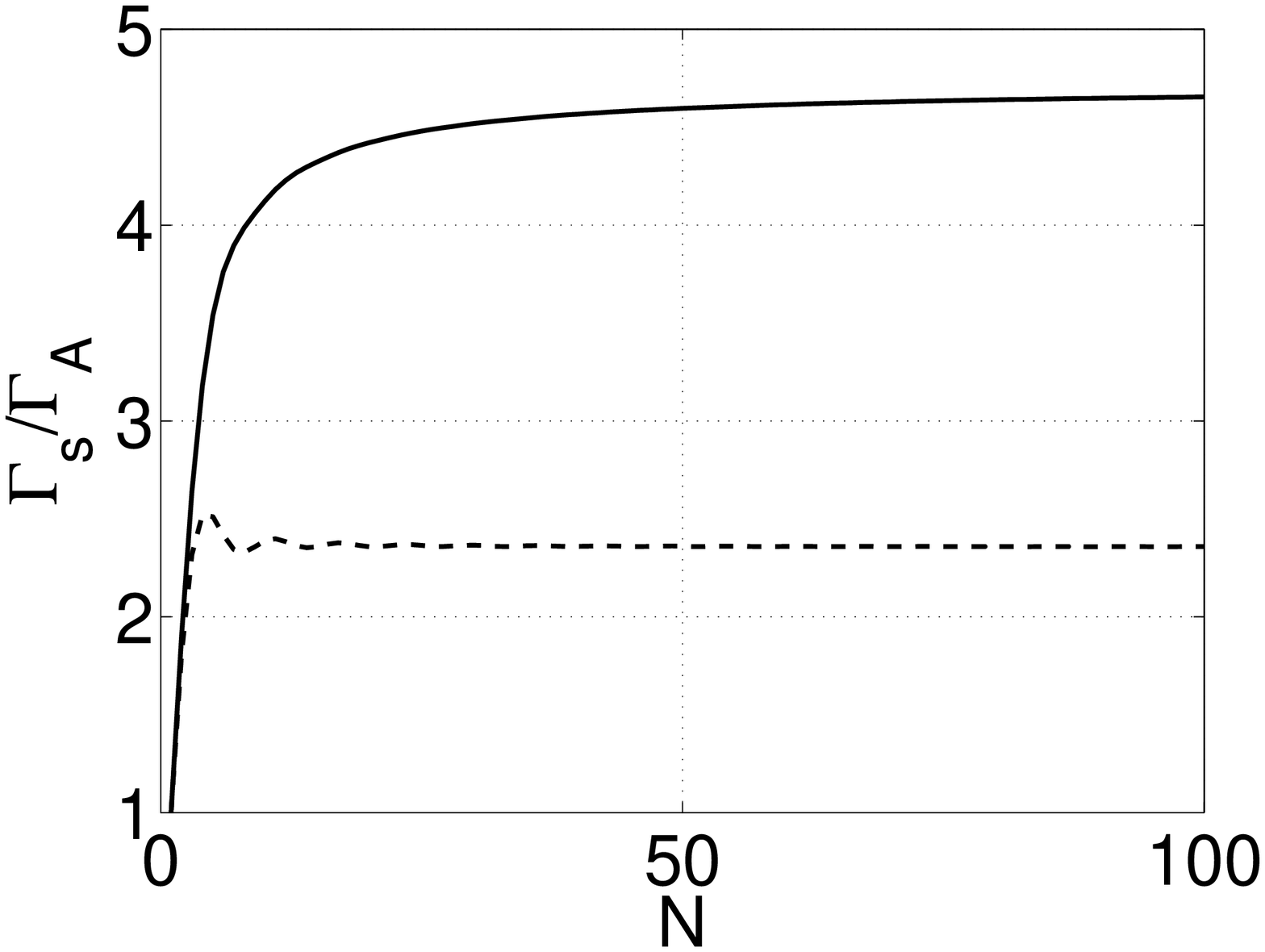}}
\caption{The symmetric scaled damping rate $\Gamma/\Gamma_A$ vs. $N$, for $q_Aa=1$. The full line is for $\varphi=0^{\circ}$, and the dashed line for $\varphi=90^{\circ}$.}
\end{figure}

\begin{figure}[h!]
\centerline{\epsfxsize=8cm \epsfbox{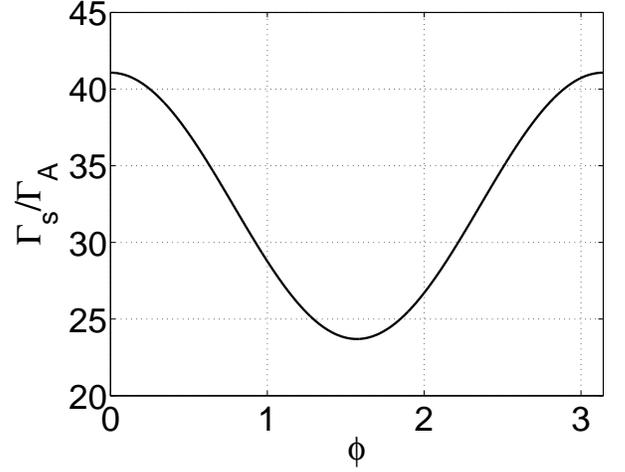}}
\caption{The symmetric scaled damping rate $\Gamma/\Gamma_A$ vs. $\varphi$, for $N=100$ at $q_Aa=0.1$.}
\end{figure}

\subsection{General Collective Excitation}

Here we consider the case of a single excitation but for a general collective state, which is given by
\begin{equation}
|i\rangle=\frac{1}{\sqrt{N}}\sum_iC_i|g_1,\cdots,e_i,\cdots,g_N\rangle,
\end{equation}
where $C_i=\pm 1$, and $\sum_iC_i^2=N$. Equation (\ref{Gamma}) reads
\begin{eqnarray}
\Gamma&=&\frac{\mu^2q_A^2}{8\pi\epsilon_0 \hbar aN}\int_{-q_Aa}^{+q_Aa}dy\left|\sum_{n=1}^NC_ne^{-iny}\right|^2 \nonumber \\
&\times&\left[\left(1+\cos^2\varphi\right)-\frac{y^2}{(q_Aa)^2}\left(3\cos^2\varphi-1\right)\right].
\end{eqnarray}
We use
\begin{equation}
\left|\sum_{n=1}^NC_ne^{-iny}\right|^2=N+\sum_{n\neq m=1}^NC_nC_me^{-i(n-m)y},
\end{equation}
where $C_nC_m=\pm 1$. After integration over $y$, we get
\begin{equation}
\Gamma(N)=\Gamma_A\left\{1+\frac{2}{N}\sum_{n<m=1}^NC_nC_m\ F[q_Aa(n-m)]\right\}.
\end{equation}

Here we present the results for two examples. For $N=2$ we have the symmetric state
\begin{equation}
|i\rangle_s=\frac{|e_1,g_2\rangle+|g_1,e_2\rangle}{\sqrt{2}},
\end{equation}
with the damping rate
\begin{equation}
\Gamma_s=\Gamma_A\left\{1+F(q_Aa)\right\},
\end{equation}
and the antisymmetric state
\begin{equation}
|i\rangle_a=\frac{|e_1,g_2\rangle-|g_1,e_2\rangle}{\sqrt{2}},
\end{equation}
with the damping rate
\begin{equation}
\Gamma_a=\Gamma_A\left\{1-F(q_Aa)\right\}.
\end{equation}
The results for $N=2$ agree with the known results \cite{Ficek}. In figure $(11)$ we plot $\Gamma/\Gamma_A$ for the symmetric state of $N=2$ as a function of $q_Aa$, for the polarizations $\varphi=0^{\circ}$ and $\varphi=90^{\circ}$. In figure $(12)$ the plot is for the antisymmetric state.

\begin{figure}[h!]
\centerline{\epsfxsize=8cm \epsfbox{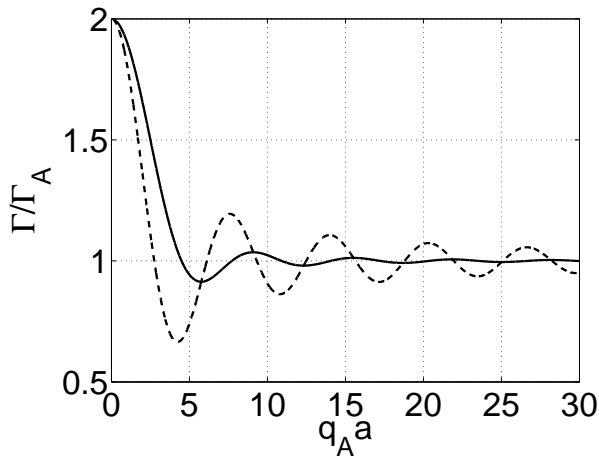}}
\caption{The scaled damping $\Gamma/\Gamma_A$ vs. $q_Aa$, for $N=2$ with the symmetric state. The full line is for $\varphi=0^{\circ}$, and the dashed line for $\varphi=90^{\circ}$.}
\end{figure}

\begin{figure}[h!]
\centerline{\epsfxsize=8cm \epsfbox{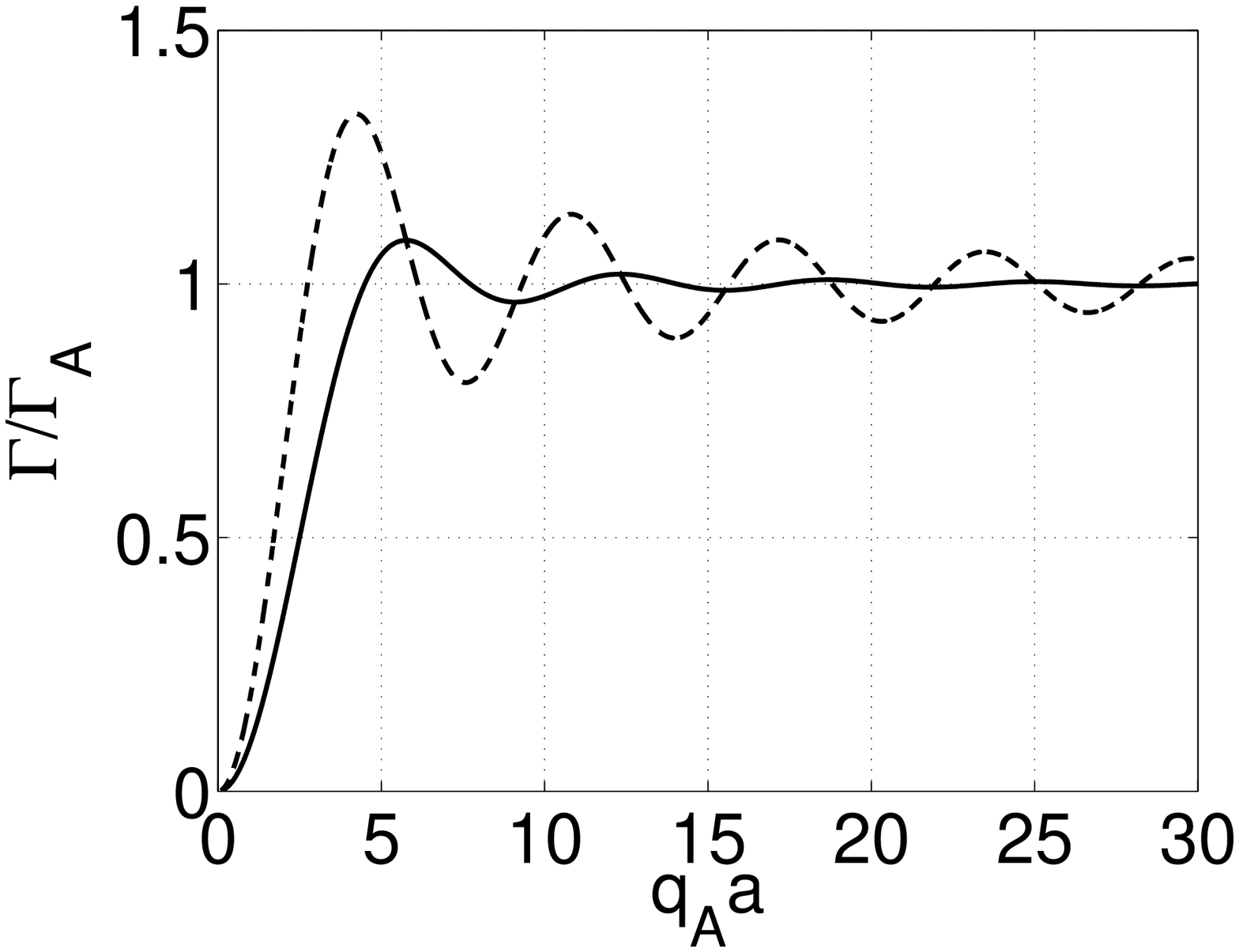}}
\caption{The scaled damping $\Gamma/\Gamma_A$ vs. $q_Aa$, for $N=2$ with the antisymmetric state. The full line is for $\varphi=0^{\circ}$, and the dashed line for $\varphi=90^{\circ}$.}
\end{figure}

For $N=3$, for the symmetric state
\begin{equation}
|i\rangle_s=\frac{|e_1,g_2,g_3\rangle+|g_1,e_2,g_3\rangle+|g_1,g_2,e_3\rangle}{\sqrt{3}},
\end{equation}
we get
\begin{equation}
\Gamma_s=\Gamma_A\left\{1+\frac{2}{3}\left[2F(q_Aa)+F(2q_Aa)\right]\right\}.
\end{equation}
For the antisymmetric state
\begin{equation}
|i\rangle_a=\frac{|e_1,g_2,g_3\rangle-|g_1,e_2,g_3\rangle+|g_1,g_2,e_3\rangle}{\sqrt{3}},
\end{equation}
we get
\begin{equation}
\Gamma_a=\Gamma_A\left\{1-\frac{2}{3}\left[2F(q_Aa)-F(2q_Aa)\right]\right\}.
\end{equation}
In the limit of $q_Aa\ll1$ we have $F\simeq 1$, then for the symmetric state we get $\Gamma\approx 3\Gamma_A$, and for the antisymmetric one we get $\Gamma\approx\Gamma_A/3$. In the limit of $q_Aa\gg1$ we have $F\simeq 0$, then for the symmetric and antisymmetric states we get $\Gamma\approx\Gamma_A$.

In figures $(13)$ we plot $\Gamma/\Gamma_A$ for the symmetric state of $N=3$ as a function of $q_Aa$, for the polarizations $\varphi=0^{\circ}$ and $\varphi=90^{\circ}$. In figure $(14)$ the plot is for the antisymmetric state.

\begin{figure}[h!]
\centerline{\epsfxsize=8cm \epsfbox{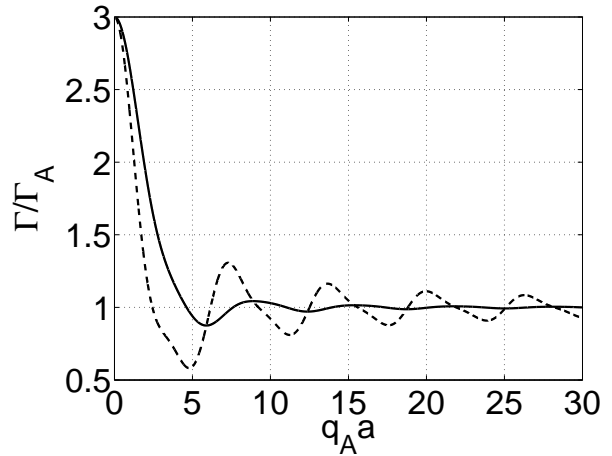}}
\caption{The scaled damping $\Gamma/\Gamma_A$ vs. $q_Aa$, of the symmetric state for $N=3$. The full line is for $\varphi=0^{\circ}$, and the dashed line for $\varphi=90^{\circ}$.}
\end{figure}

\begin{figure}[h!]
\centerline{\epsfxsize=8cm \epsfbox{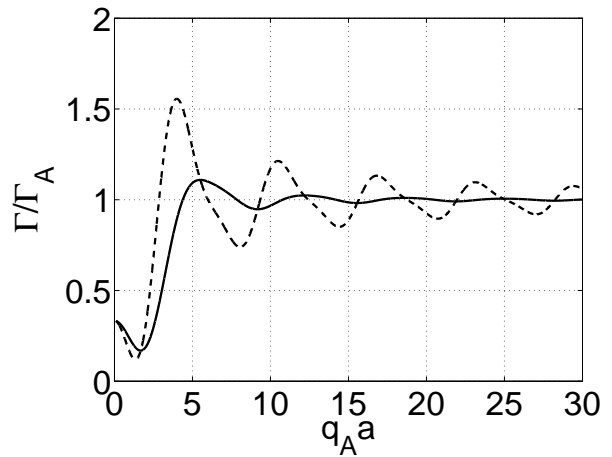}}
\caption{The scaled damping $\Gamma/\Gamma_A$ vs. $q_Aa$, of the antisymmetric state for $N=3$. The full line is for $\varphi=0^{\circ}$, and the dashed line for $\varphi=90^{\circ}$.}
\end{figure}

\section{Collective Excitation Emission Pattern}

Here we calculate the emission pattern of a collective state in a chain of $N$ atoms separated by a distance $a$. The transition dipole of each atom is $\mbox{\boldmath$\mu$}=\mu(\sin\varphi,0,\cos\varphi)$, at positions ${\bf R}_n=(0,0,R_n)$. For simplicity the observation point is taken to be at ${\bf r}=(x,0,0)$, as seen in figure $(15)$. We concentrate here in the limit of $q_Aa>1$ where the atoms can be treated independently. The other limit of  $q_Aa<1$ investigated by us in previous work \cite{ZoubiH}. The positive electric field operator of the atom $(n)$, in the far zone field where $x\gg\lambda_A$, is given by \cite{Loudon}
\begin{equation}
\hat{\bf E}^{(+)}_n({\bf r},t)=\frac{\mu q_A^2}{4\pi\epsilon_0}\frac{\sin\phi_n}{|{\bf r}-{\bf R}_n|}B\left(t-\frac{|{\bf r}-{\bf R}_n|}{c}\right)\hat{\bf e}_n,
\end{equation}
where $\phi_n$ is the angle between $\mbox{\boldmath$\mu$}$ and ${\bf r}-{\bf R}_n$, and the unit vector $\hat{\bf e}_n$ is defined by
\begin{equation}
\hat{\bf e}_n=\hat{\bf n}_n\times\hat{\bf y},\ \hat{\bf n}_n=\frac{{\bf r}-{\bf R}_n}{|{\bf r}-{\bf R}_n|}.
\end{equation}
We have
\begin{equation}
{\bf r}-{\bf R}_n=(x,0,-R_n),\ |{\bf r}-{\bf R}_n|^2=x^2+R_n^2,
\end{equation}
and
\begin{equation}
\phi_n=\pi-\varphi-\alpha_n,\ \tan\alpha_n=x/R_n,
\end{equation}
with
\begin{equation}
\hat{\bf n}_n=\frac{(x,0,-R_n)}{\sqrt{x^2+R_n^2}},\ \hat{\bf e}_n=\frac{(R_n,0,x)}{\sqrt{x^2+R_n^2}}.
\end{equation}

\begin{figure}[h!]
\centerline{\epsfxsize=6cm \epsfbox{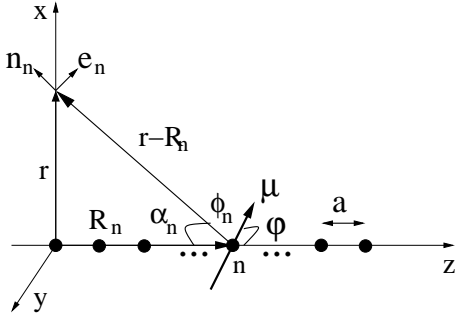}}
\caption{The observation point at ${\bf r}$ along the ${\bf x}$ axis, and the lattice is along the ${\bf z}$ axis. The angle between the transition dipole $\mbox{\boldmath$\mu$}$ and ${\bf r}-{\bf R}_n$ is $\phi_n$. The electric field direction off atom $(n)$ is $\hat{\bf e}_n$.}
\end{figure}

For the atomic transition operators we use the expectation values
\begin{eqnarray}
\langle B_i(t-t_i)\rangle&=&\langle B_i(0)\rangle e^{-i\omega_A(t-t_i)} e^{-\Gamma_A(t-t_i)/2}, \nonumber \\
\langle B_i^{\dagger}(t-t_i)B_i(t-t_i)\rangle&=&\langle B_i^{\dagger}(0)B_i(0)\rangle e^{-\Gamma_A(t-t_i)},
\end{eqnarray}
and
\begin{eqnarray}
\langle B_i^{\dagger}(t-t_i)B_j(t-t_j)\rangle&=&\langle B_i^{\dagger}(0)B_j(0)\rangle e^{-\Gamma_A\left[t-(t_i+t_j)/2\right]} \nonumber \\
&\times&e^{-i\omega_A(t_i-t_j)},
\end{eqnarray}
where $t_i=|{\bf r}-{\bf R}_i|/c$. In the limit $q_Aa>1$ the single excitation collective states decay with the single excited atom damping rate $\Gamma_A$.

The total electric field at the observation point is
\begin{equation}
\hat{\bf E}^{(+)}({\bf r},t)=\sum_i\hat{\bf E}_i^{(+)}({\bf r},t),
\end{equation}
and the intensity is
\begin{equation}
I({\bf r},t)=\frac{1}{2}\epsilon_0c\langle\hat{\bf E}^{(-)}({\bf r},t)\mbox{\boldmath$\cdot$}\hat{\bf E}^{(+)}({\bf r},t)\rangle.
\end{equation}
Explicitly we can write
\begin{equation}
I({\bf r},t)=\sum_iI_i({\bf r},t)+\sum_{i\neq j}G_{ij}({\bf r},t),
\end{equation}
where the $i$-th intensity is
\begin{equation}
I_i({\bf r},t)=\frac{1}{2}\epsilon_0c\langle\hat{\bf E}_i^{(-)}({\bf r},t)\hat{\bf E}_i^{(+)}({\bf r},t)\rangle,
\end{equation}
and the correlation function is
\begin{equation}
G_{ij}({\bf r},t)=\frac{1}{2}\epsilon_0c\langle\hat{\bf E}_i^{(-)}({\bf r},t)\mbox{\boldmath$\cdot$}\hat{\bf E}_j^{(+)}({\bf r},t)\rangle.
\end{equation}
We get
\begin{equation}
I_i({\bf r},t)=\frac{\mu^2 \omega_A^4}{32\pi^2\epsilon_0c^3}\frac{\sin^2\phi_i}{|{\bf r}-{\bf R}_i|^2}\langle B_i^{\dagger}(0)B_i(0)\rangle e^{-\Gamma_A(t-t_i)},
\end{equation}
and
\begin{eqnarray}
G_{ij}({\bf r},t)&=&\frac{\mu^2 \omega_A^4}{32\pi^2\epsilon_0c^3}e^{-\Gamma_A\left[t-(t_i+t_j)/2\right]}e^{-i\omega_A(t_i-t_j)} \nonumber \\
&\times& \langle B_i^{\dagger}(0)B_j(0)\rangle\frac{\sin\phi_i}{|{\bf r}-{\bf R}_i|}\frac{\sin\phi_j}{|{\bf r}-{\bf R}_j|}\left(\hat{\bf n}_i\mbox{\boldmath$\cdot$}\hat{\bf n}_j\right). \nonumber \\
\end{eqnarray}

\subsection{Two-Atoms Chain}

We present the results for the simple case of two atoms. One atom is located at the origin ${\bf R}_1=(0,0,0)$, and the second at ${\bf R}_2=(0,0,a)$. The observation point is at ${\bf r}=(x,0,0)$, where ${\bf r}-{\bf R}_1=(x,0,0)$, and ${\bf r}-{\bf R}_2=(x,0,-a)$, with $|{\bf r}-{\bf R}_1|=x$, and $|{\bf r}-{\bf R}_2|=\sqrt{x^2+a^2}$. We have the angles $\phi_1=\frac{\pi}{2}-\varphi$, and $\phi_2=\pi-\varphi-\alpha$, where $\tan\alpha=x/a$. We get the times $t_1=x/c$, and $t_2=\sqrt{x^2+a^2}/c$. Also we have the unit vectors $\hat{\bf e}_1=(0,0,1)$, and $\hat{\bf e}_2=\frac{(a,0,x)}{\sqrt{x^2+a^2}}$, then $\hat{\bf n}_1=(1,0,0)$, and $\hat{\bf n}_2=\frac{(x,0,-a)}{\sqrt{x^2+a^2}}$, hence $\left(\hat{\bf n}_1\mbox{\boldmath$\cdot$}\hat{\bf n}_2\right)=\frac{x}{\sqrt{x^2+a^2}}$. We obtain
\begin{eqnarray}
I_1({\bf r},t)&=&\frac{\mu^2 \omega_A^4}{32\pi^2\epsilon_0c^3}\ \frac{\sin^2\phi_1}{x^2}\ \langle B_1^{\dagger}(0)B_1(0)\rangle\ e^{-\Gamma_A\left(t-\frac{x}{c}\right)}, \nonumber \\
I_2({\bf r},t)&=&\frac{\mu^2 \omega_A^4}{32\pi^2\epsilon_0c^3}\ \frac{\sin^2\phi_2}{x^2+a^2}\ \langle B_2^{\dagger}(0)B_2(0)\rangle \nonumber \\
&\times&e^{-\Gamma_A\left(t-\frac{\sqrt{x^2+a^2}}{c}\right)},
\end{eqnarray}
and
\begin{eqnarray}
G_{12}({\bf r},t)&=&\frac{\mu^2 \omega_A^4}{32\pi^2\epsilon_0c^3}\ \frac{\sin\phi_1\sin\phi_2}{x^2+a^2}\ \langle B_1^{\dagger}(0)B_2(0)\rangle \nonumber \\
&\times&e^{-\Gamma_A\left[t-\left(\frac{x+\sqrt{x^2+a^2}}{2c}\right)\right]}\ e^{-i\omega_A\left(\frac{x-\sqrt{x^2+a^2}}{c}\right)}, \nonumber \\
G_{21}({\bf r},t)&=&\frac{\mu^2 \omega_A^4}{32\pi^2\epsilon_0c^3}\ \frac{\sin\phi_1\sin\phi_2}{x^2+a^2}\ \langle B_2^{\dagger}(0)B_1(0)\rangle \nonumber \\
&\times&e^{-\Gamma_A\left[t-\left(\frac{x+\sqrt{x^2+a^2}}{2c}\right)\right]}\ e^{i\omega_A\left(\frac{x-\sqrt{x^2+a^2}}{c}\right)}.
\end{eqnarray}
Now we consider the two initial states of symmetric and antisymmetric collective states.

For the symmetric collective state
\begin{equation}
|i\rangle=\frac{|e_1,g_2\rangle+|g_1,e_2\rangle}{\sqrt{2}},
\end{equation}
we have
\begin{eqnarray}
\langle B_1^{\dagger}(0)B_1(0)\rangle&=&\langle B_2^{\dagger}(0)B_2(0)\rangle \nonumber \\
=\langle B_1^{\dagger}(0)B_2(0)\rangle&=&\langle B_2^{\dagger}(0)B_1(0)\rangle=\frac{1}{2},
\end{eqnarray}
then we get
\begin{eqnarray}
I({\bf r},t)&=&\frac{I_0(x)}{4}\left\{\sin^2\phi_1\ e^{-\Gamma_A\left(t-\frac{x}{c}\right)}+\frac{x^2\sin^2\phi_2}{x^2+a^2}\right. \nonumber \\
&\times&\left.e^{-\Gamma_A\left(t-\frac{\sqrt{x^2+a^2}}{c}\right)}\right. \nonumber \\
&+&\left.\frac{x^2\sin\phi_1\sin\phi_2}{x^2+a^2}\ 2\cos\left[\omega_A\left(\frac{x-\sqrt{x^2+a^2}}{c}\right)\right]\right. \nonumber \\
&\times&\left.e^{-\Gamma_A\left[t-\left(\frac{x+\sqrt{x^2+a^2}}{2c}\right)\right]}\right\},
\end{eqnarray}
where we defined the intensity
\begin{equation}
I_0(x)=\frac{\mu^2 \omega_A^4}{16\pi^2\epsilon_0c^3x^2}.
\end{equation}

In figures $(16-18)$ we plot the relative intensity $I({\bf r},t)/I_0(x)$ as a function of $a$ for the angles $\varphi=0^{\circ}$, $\varphi=45^{\circ}$ and $\varphi=90^{\circ}$, at the observation point $x=10^{6}\ \AA$ at the moment $t=2x/c$. We use $E_A=1\ eV$, $\mu=1\ e\AA$ and $\Gamma_A=10^{8}\ Hz$. 

For small $a$ the relative intensity is maximum for the polarization angle $\varphi=0^{\circ}$ and decreases for larger angles. It is half for $\varphi=45^{\circ}$, and becomes zero for $\varphi=90^{\circ}$. The maximum of the relative intensity moves into larger $a$ with increasing the angle $\varphi$. The relative intensity oscillates in changing $a$ and tend to a finite value for large $a$. Interesting case is for $\varphi=90^{\circ}$, where the intensity is zero for small $a$ and increases with increasing $a$ till it reach a maximum at $a=10^{6}\ \AA$ (for the given numbers), and decreases back towards a finite value for larger $a$.

In the limit of $x\gg a$, where $\sqrt{x^2+a^2}\sim x+\frac{a^2}{2x}$, and as $\phi\approx\frac{\pi}{2}-\varphi$, we can write
\begin{eqnarray}
I({\bf r},t)&\simeq&\frac{\mu^2 \omega_A^4}{64\pi^2\epsilon_0c^3x^2}\ e^{-\Gamma_A\left(t-\frac{x}{c}\right)}\cos^2\varphi \nonumber \\
&\times&\left\{1+e^{\Gamma_A\frac{{a^2}}{2cx}}+2\cos\left(\omega_A\frac{a^2}{2cx}\right)\ e^{\Gamma_A\frac{a^2}{4cx}}\right\}.
\end{eqnarray}

\begin{figure}[h!]
\centerline{\epsfxsize=8cm \epsfbox{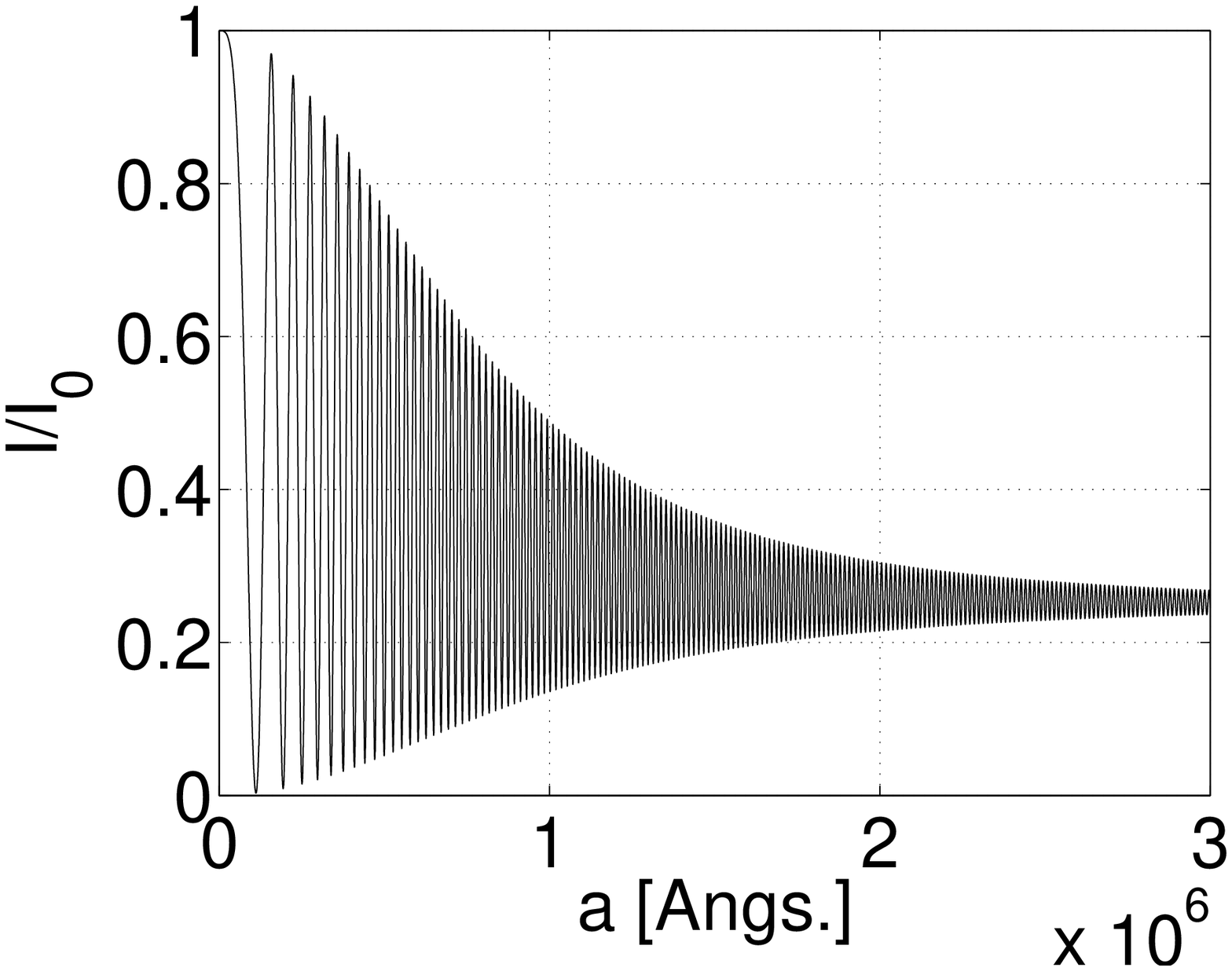}}
\caption{The symmetric state scaled intensity $I({\bf r},t)/I_0(x)$ vs. $a$, for $\varphi=0^{\circ}$ at $x=10^{6}\ \AA$ and $t=2x/c$.}
\end{figure}

\begin{figure}[h!]
\centerline{\epsfxsize=8cm \epsfbox{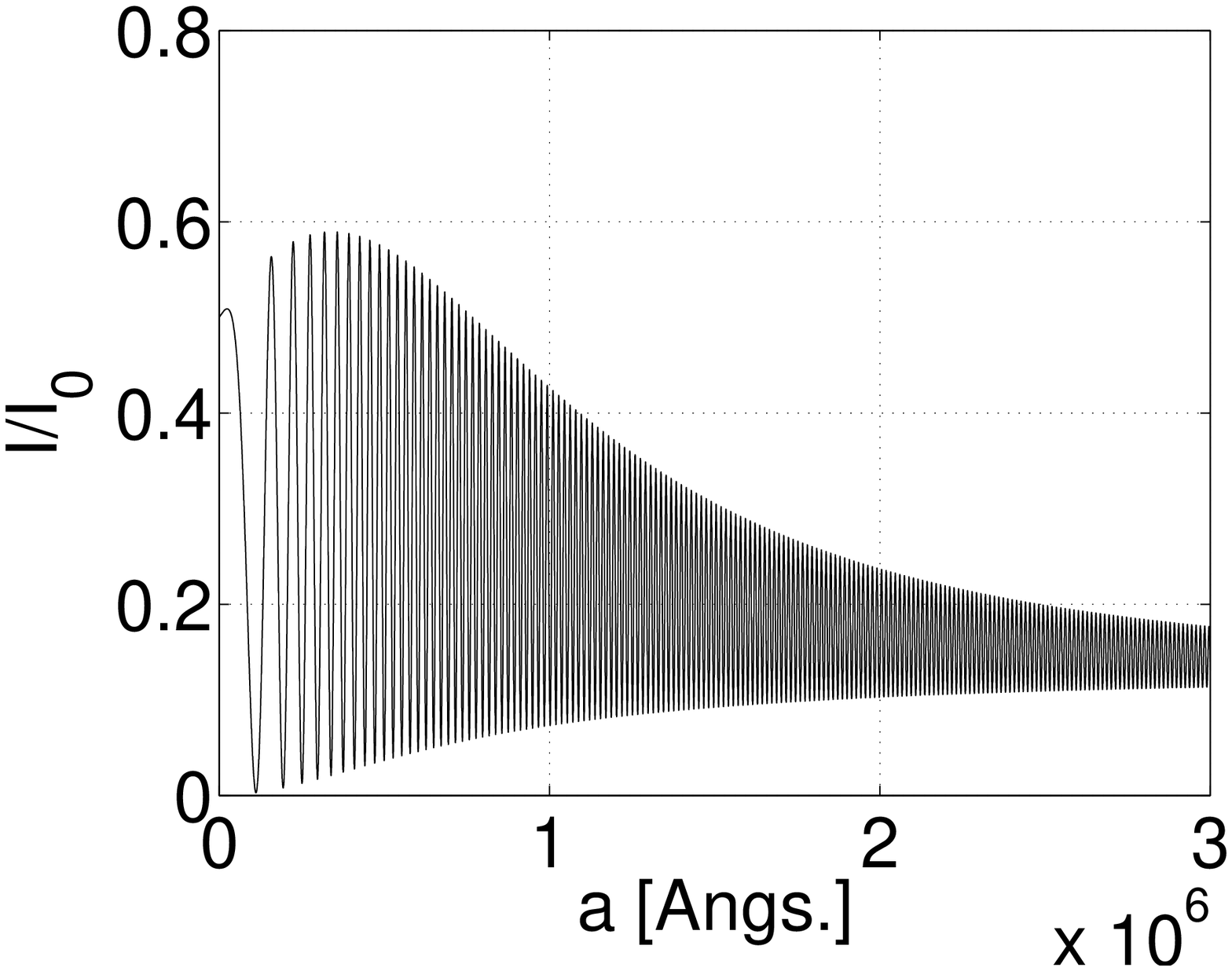}}
\caption{The symmetric state scaled intensity $I({\bf r},t)/I_0(x)$ vs. $a$, for $\varphi=45^{\circ}$ at $x=10^{6}\ \AA$ and $t=2x/c$.}
\end{figure}

\begin{figure}[h!]
\centerline{\epsfxsize=8cm \epsfbox{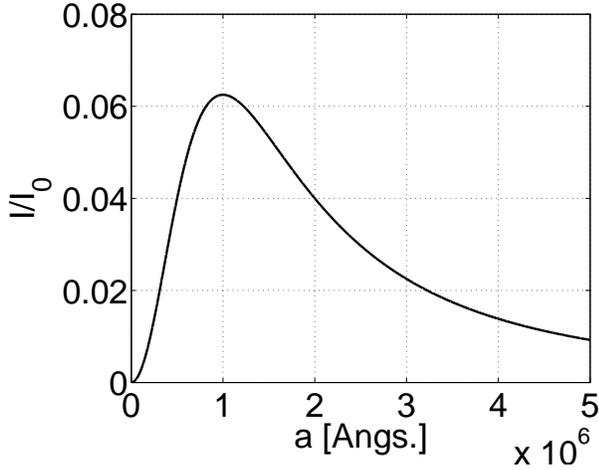}}
\caption{The symmetric or antisymmetric state scaled intensity $I({\bf r},t)/I_0(x)$ vs. $a$, for $\varphi=90^{\circ}$ at $x=10^{6}\ \AA$ and $t=2x/c$.}
\end{figure}

For the antisymmetric collective state
\begin{equation}
|i\rangle=\frac{|e_1,g_2\rangle-|g_1,e_2\rangle}{\sqrt{2}},
\end{equation}
we have
\begin{eqnarray}
\langle B_1^{\dagger}(0)B_1(0)\rangle&=&\langle B_2^{\dagger}(0)B_2(0)\rangle=\frac{1}{2}, \nonumber \\
\langle B_1^{\dagger}(0)B_2(0)\rangle&=&\langle B_2^{\dagger}(0)B_1(0)\rangle=-\frac{1}{2},
\end{eqnarray}
then we can write
\begin{eqnarray}
I({\bf r},t)&=&\frac{I_0(x)}{4}\left\{\sin^2\phi_1\ e^{-\Gamma_A\left(t-\frac{x}{c}\right)}+\frac{x^2\sin^2\phi_2}{x^2+a^2}\right. \nonumber \\
&\times&\left.e^{-\Gamma_A\left(t-\frac{\sqrt{x^2+a^2}}{c}\right)}\right. \nonumber \\
&-&\left.\frac{x^2\sin\phi_1\sin\phi_2}{x^2+a^2}\ 2\cos\left[\omega_A\left(\frac{x-\sqrt{x^2+a^2}}{c}\right)\right]\right. \nonumber \\
&\times&\left.e^{-\Gamma_A\left[t-\left(\frac{x+\sqrt{x^2+a^2}}{2c}\right)\right]}\right\}.
\end{eqnarray}
In figures $(19-20)$ we plot the relative intensity $I({\bf r},t)/I_0(x)$ as a function of $a$ for the angles $\varphi=0^{\circ}$ and $\varphi=45^{\circ}$. The case of $\varphi=90^{\circ}$ is the same as in figure $(23)$. As before, the observation point is at $x=10^{6}\ \AA$ at the moment $t=2x/c$, with the other previous numbers. The results are similar to the symmetric ones except from the case of small $a$ where the relative intensity tends to zero as expected.

In the limit of $x\gg a$, as $\phi\approx\frac{\pi}{2}-\varphi$, we can write
\begin{eqnarray}
I({\bf r},t)&\simeq&\frac{\mu^2 \omega_A^4}{64\pi^2\epsilon_0c^3x^2}\ e^{-\Gamma_A\left(t-\frac{x}{c}\right)}\cos^2\varphi \nonumber \\
&\times&\left\{1+e^{\Gamma_A\frac{{a^2}}{2cx}}-2\cos\left(\omega_A\frac{a^2}{2cx}\right)\ e^{\Gamma_A\frac{a^2}{4cx}}\right\}.
\end{eqnarray}

\begin{figure}[h!]
\centerline{\epsfxsize=8cm \epsfbox{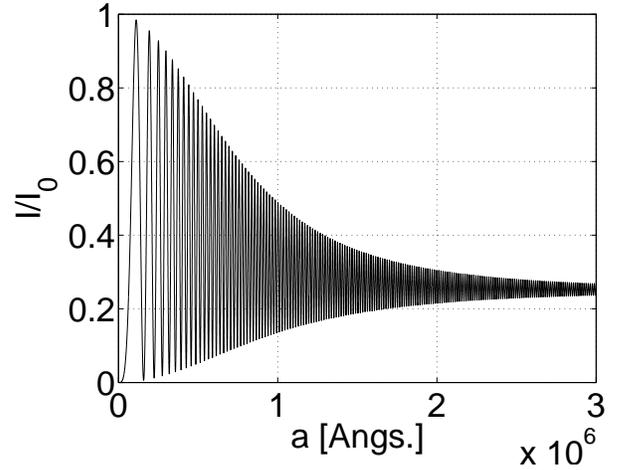}}
\caption{The antisymmetric state scaled intensity $I({\bf r},t)/I_0(x)$ vs. $a$, for $\varphi=0^{\circ}$ at $x=10^{6}\ \AA$ and $t=2x/c$.}
\end{figure}

\begin{figure}[h!]
\centerline{\epsfxsize=8cm \epsfbox{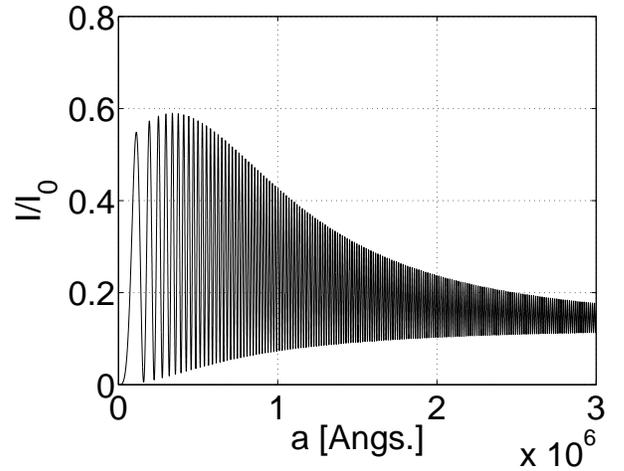}}
\caption{The antisymmetric state scaled intensity $I({\bf r},t)/I_0(x)$ vs. $a$, for $\varphi=45^{\circ}$ at $x=10^{6}\ \AA$ and $t=2x/c$.}
\end{figure}

\section{Summary}

In the present paper we investigated optical properties of a one dimensional atomic chain, in which the lattice constant can range from a few angstroms up to thousands of angstroms. Namely, the lattice constant can change from being smaller than the atomic transition wavelength up to much larger one. In the limit of lattice constant smaller than the atomic transition wavelength the electrostatic interactions are applicable, which found useful for most of the typical experiments on optical lattice ultracold atoms. In our previous work we limited the discussion to electrostatic interactions, where we considered only resonance dipole-dipole interactions, and which is justified in the present work. For small lattice constant the electrostatic interactions are responsible for the formation of excitons, where we did extensive study in this regime with emphasize on the exciton life times. For large lattice constant the inclusion of radiative corrections are necessary, which is the main issue in the present paper.

For large lattice constant the radiative corrections are included, and in this regime we found that the coupling parameter for the energy transfer among even the nearest neighbor atom sites is smaller than a single excited atom damping rate. Hence, the energy transfer is not favorable, and we treated the atoms as independently setting on the lattice sites. Then, we calculated the damping rates of different collective electronic excitations in including the radiative corrections by considering the effect of the existence of all the other atom sites, despite their large distances from the excited atom. Big attention we gave for the most symmetric state, where we emphasized the dependence of its damping rate on the number of atoms for different lattice constant. We found the symmetric damping rate to behave linearly at small atom numbers and saturate at large numbers. The damping rate of symmetric and antisymmetric collective states tend to that of a single excited atom with oscillations due to the radiative effect through the exchange of virtual photons. The differences between damping rates of collective states appear for small lattice constant, in which the symmetric states have superradiant damping rate, that is $N$ time the single excited atom rate. Here, part of the antisymmetric states become dark with zero damping rate, and other part is metastable with a fraction of the single excited atom damping rate. Moreover we calculated the emission pattern off a chain of atoms with a large lattice constant in which the atoms can be considered independently. The emission intensities off two atoms with symmetric and antisymmetric states are presented as a function of the interatomic distance.

The results of the present paper are illustrated in terms of optical lattice ultracold atoms, but they are general and can be adopted for any chain of optically active material. For example, chains of semiconductor quantum dots fit exactly in the regime of large lattice constant, where radiative corrections are unavoidable, and the life times of their collective states can be treated according to the present paper. Other system that exploits the regime of the present paper is a lattice of large organic molecules sitting on a matrix with a given large lattice constant, the collective damping rate and emission pattern are expected to behave according to our present results.

\ 

The author acknowledge very fruitful discussions with Helmut Ritsch. The work was supported by the Austrian Science Funds (FWF) via the project (P21101), and by the DARPA QuASAR program.

\end{document}